**A.I. Glushchenko, Dr. Sc. (aiglush@ipu.ru),**
**K.A. Lastochkin (lastconst@ipu.ru)**
**(V.A. Trapeznikov Institute of Control Sciences of Russian Academy of Sciences, Moscow),**
**V.A. Petrov, C. Sc. (petrov.va@misis.ru),**
**(Stary Oskol technological institute n.a. A.A. Ugarov (branch) National University of Science and Technology "MISIS", Stary Oskol)**

# EXPONENTIALLY STABLE ADAPTIVE CONTROL. PART I. TIME-INVARIANT PLANTS[1]

**Abstract.** In this research we consider linear time-invariant plants and assume that the regressor finite excitation requirement is met. In such case, a new law to adjust the controller parameters, which ensures the exponential stability of the classical dynamic model of the tracking error under the condition that its states are not included in such a law, is proposed in this study. In addition, it also relaxes a number of classical assumptions and requirements of the adaptive control theory, i.e. the necessity to know the sign/value of the plant high-frequency gain, the need of experimentally based choice of the proposed law adaptive gain value, the requirement to the tracking error transfer function to be strictly positive real considering the output feedback control. The applicability of the proposed law to the problems of adaptive state and output feedback control is shown. The advantages of the proposed method over the well-known ones are demonstrated.

## 1. Introduction

In the adaptive control literature, a common form of representation of a plant with parametric uncertainty is a dynamic error equation [1,2]:

$$\dot{e}_{ref} = A_{ref}e_{ref} + B\tilde{\theta}^T\omega, \ e_{ref}(0) = e_{0ref}, \quad (1.1)$$

where $e_{ref} \in R^q$ is a vector of tracking errors calculated as the difference between a plant and a reference model states, $e_{ref}(0) \in R^q$ is the initial conditions vector, $A_{ref} \in R^{q \times q}$ is a Hurwitz reference model state matrix, $B \in R^q$ is a plant input vector (high frequency gain), $\tilde{\theta} = \hat{\theta} - \theta \in R^m$ is a parametric error calculated as the difference between the adjustable and ideal (unknown but time-invariant $\dot{\theta} = 0$) control law parameters, $\omega \in R^m$ is a measurable regressor. The matrix $A_{ref}$ is known. The pair ($A_{ref}, B$) is fully controllable.

In case, when only the output tracking error is measurable, the equation (1.1) takes the form:

$$\dot{e}_{ref} = A_{ref}e_{ref} + B\tilde{\theta}^T\omega, \ e_{ref}(0) = e_{0ref},$$
$$\varepsilon = c^T e_{ref}, \quad (1.2)$$

$c \in R^q$ is the vector to form the measurable error $\varepsilon$.

Usually, the error equation (1.1) is a result of the parametrization applied to the state feedback adaptive control problems, whereas the model (1.2) is obtained for the output feedback adaptive control problems (see [1-3] for details).

---

[1] This research was financially supported in part by Grants Council of the President of the Russian Federation (project MD-1787.2022.4).

It is well known [1-3] that, if the all the components of the tracking error vector $e_{ref}$ are measurable, to provide the asymptotic stability of the model (1.1) it is enough to choose the adaptive law of the controller parameters in the following form:

$$\dot{\hat{\theta}} = -\Gamma \omega e_{ref}^T P B, \quad (1.3)$$

where $\Gamma$ is an adaptive gain of an appropriate dimension, $P \in R^{q \times q}$ is the Lyapunov equation solution: $A_{ref} P + P A_{ref}^T = -Q$, in which $Q \in R^{q \times q}$ is a positively definite matrix.

It is also known that, if a transfer function $c^T (pI - A_{ref})^{-1} B$ is strictly positive real (SPR), then there exists the solution $P$ of the following system of equations [1-3]:

$$\begin{aligned} A_{ref} P + P A_{ref}^T &= -Q, \\ PB &= c^T. \end{aligned} \quad (1.4)$$

In such case the asymptotic stability of the error model (1.2) is provided with the help of the following adaptive law:

$$\dot{\hat{\theta}} = -\Gamma \omega \varepsilon. \quad (1.5)$$

If, additionally, the regressor persistent excitation (PE) requirement $\omega \in PE$ is met, then the adaptive laws (1.3) and (1.5) ensure [1,2] exponential stability of the augmented tracking error $\xi = \begin{bmatrix} e_{ref}^T & \tilde{\theta}^T \end{bmatrix}^T$.

The main drawbacks of the adaptive laws (1.3) and (1.5) are: (a) that the exponential stability of the error $\xi$ is provided only if the restrictive condition $\omega \in PE$ is met, and (b) the poor quality of the transients with respect to the tracking errors $e_{ref}$, $\varepsilon$ and the adjustable parameters $\hat{\theta}$. To overcome these disadvantages, various modifications [4-12] of the adaptive laws (1.3) and (1.5) have been proposed in the literature to relax the PE condition and/or to improve the transient quality of the tracking errors $e_{ref}$, $\varepsilon$ and the adjustable parameters $\hat{\theta}$.

Moreover, as far as the output feedback control problem is concerned, the adaptive law (1.5) can be implemented if and only if the transfer function $c^T (pI - A_{ref})^{-1} B$ is SPR [1-3,13]. To avoid the SPR requirement, the following approaches have been developed [3,13,14]: augmented error method [15], high order tuners [16], shunt compensator [17], iterative synthesis procedures [18] and their various modifications [3]. However, most of them are noise-sensitive, tedious, and rather complex to be implemented in practice [14]. The higher the relative degree of the plant, the more evident the latter two problems become.

In this research, as far as the problems of both state and output feedback adaptive control are concerned, in order to overcome the feasibility conditions of law (1.5) and improve the quality of the controller parameters adjustment process, a new adaptive law is proposed, which has the following properties.

1) It does not include the states of the error equation (1.1).
2) It is equally applicable to error models (1.1) and (1.2).
3) If the regressor finite excitation requirement is met, it provides exponential stability of the error $\xi$ and elementwise monotonicity of transients of the controller adjustable parameters $\hat{\theta}$.
4) It solves several classical problems of the adaptive control theory: the requirement to know the elements or sign of the plant input vector $B$, the need to choose the adaptive gain matrix $\Gamma$ manually.

The remainder of the paper is organized as follows. In Section 2 the existence theorem of a new adaptive law to adjust the controller parameters $\hat{\theta}$, which ensures exponential stability of the error $\xi$ under the condition that the regressor is finitely exciting, is formulated, the simple algorithms of its implementation to solve the problems of the state and output feedback adaptive control are proposed. Section 3 presents the results of numerical experiments.

The definition of the regressor finite excitation and the corollary of the Kalman-Yakubovich-Popov Lemma will be used in the proofs of Theorem and Propositions.

**Definition 1:** *The regressor $\omega$ is finitely exciting ($\omega \in$ FE) over the time range $\left[t_r^+;\ t_e\right]$ if there exist $t_r^+ \geq 0$, $t_e \geq t_r^+$ and $\alpha$ such that the following inequality holds:*

$$\int_{t_r^+}^{t_e} \omega(\tau)\omega^T(\tau)d\tau \geq \alpha I, \qquad (1.6)$$

*where $\alpha > 0$ is the excitation level, I is a unity matrix.*

**Corollary.** *A scalar transfer function $H(s) = d$, where $d > 0$, is strictly positive real if and only if there exist a Hurwitz matrix A, a matrix $P = P^T > 0$, a vector $q$, a vector B and a constant $\mu > 0$, such that* [1-3]:

$$\begin{aligned} A^T P + PA &= -qq^T - \mu P, \\ PB &= \sqrt{2d}\, q. \end{aligned} \qquad (1.7)$$

## 2. Main result

Let it be assumed that there exists a linear-regression-based model of the following form for an unknown parameters vector $\theta$ from (1.1) and (1.2):

$$\Upsilon = \Omega(\Delta)\theta, \qquad (2.1)$$

where $\Omega \in R$, $\Upsilon \in R^m$, $\Delta \in R$ are the measurable regressor, function and regressor argument respectively, which do not depend from the state vector $e_{ref}$ of the error equation (1.1).

Then, on the basis of regression (2.1), the existence theorem of the adaptive law to adjust the parameters $\hat{\theta}$, which does not use the error equation state vector (1.1) and provides exponential stability of the error $\xi$, can be formulated and proved.

**Theorem**. *Considering that $\Delta \in$ FE, if the regressor $\Omega$ satisfies the following requirements:*

a) $\forall t \geq t_r^+\ \Omega(t) \in L_\infty,\ \Omega(t) \geq 0;$

б) $\forall t \geq t_e\ 0 < \Omega_{LB} \leq \Omega(t) \leq \Omega_{UB},$

*then there is an adaptive law of the form:*

$$\dot{\hat{\theta}} = \dot{\tilde{\theta}} = -\gamma\Omega\big(\Omega\hat{\theta} - \Omega\theta\big) = -\gamma\Omega^2\tilde{\theta},$$

$$\gamma = \begin{cases} 1, & \text{if } \Omega = 0, \\ \dfrac{\gamma_0 \lambda_{\max}\left(\omega\omega^T\right) + \gamma_1}{\Omega^2} & \text{otherwise,} \end{cases} \qquad (2.2)$$

*which provides the following properties*:

1) $\forall t_a \geq t_b\ \left|\tilde{\theta}_i(t_a)\right| \leq \left|\tilde{\theta}_i(t_b)\right|;$

2) $\forall t \geq t_r^+\ \xi(t) \in L_\infty;$

3) $\forall t \geq t_e$ the augmented error $\xi(t) = \begin{bmatrix} e_{ref}^{\mathrm{T}}(t) & \tilde{\theta}^{\mathrm{T}}(t) \end{bmatrix}^{\mathrm{T}}$ converges to zero exponentially with the rate, which minimal value depends directly proportional from $\gamma_0 \geq 1$ and $\gamma_1 \geq 0$.

Proof of Theorem is presented in Appendix.

According to the results of Theorem, the adaptive law (2.2) is equally applicable to the dynamic models based on both measurable state vector (1.1) and measurable output signal (1.2). Compared to the adaptive laws (1.3) and (1.5), the law (2.2):

1) provides elementwise monotonicity of the adjustable controller parameters $\hat{\theta}$;
2) ensure the exponential convergence of the error $\xi$ under the condition $\Delta \in \mathrm{FE}$, which is strictly weaker than $\omega \in \mathrm{PE}$;
3) does not require to know the sign or value of elements of the vector $B$;
4) is less sensitive to the manually chosen value of the parameter $\gamma$;
5) as far as the problem of the output feedback adaptive control is concerned, unlike known solutions [15, 16], allows one to overcome the requirement to meet the SPR condition without the sufficient increase in complexity of the control system when the value of the plant relative degree improves.

**Remark 1.** *The adaptive law (2.2) is less sensitive to the manually chosen value of the parameter $\gamma$, because, first of all, it contains a dynamic term $\gamma_0 \lambda_{\max}(\omega \omega^{\mathrm{T}})$, which does not require a manual selection. And, secondly, the minimum rate of $\xi$ error convergence to zero can be set at a required level by means of $\gamma_1$ value and independently from the value of $\Omega_{\mathrm{LB}}$.*

The fundamental practical difficulty to implement the adaptive law (2.2) is to obtain a linear regression model (2.1) with a regressor $\Omega$, which satisfies the requirements of Theorem, for each particular adaptive control problem.

In the following subsections we will show how such a regression model can be obtained for the most general classical state and output feedback adaptive control problems, which could be reduced to the error models (1.1) and (1.2) respectively.

### 2.1. State feedback control case

Let a classical problem of state feedback adaptive control of LTI SISO plants be considered [1-3]:

$$\dot{x} = Ax + Bu, \quad x(0) = x_0, \quad (2.1.1)$$

where $x \in R^n$ is a measurable state vector, $x_0 \in R^n$ is an unknown vector of initial conditions, $u \in R$ is a control signal, $A \in R^{n \times n}$ is a plant state matrix, and $B \in R^n$ is a plant input vector. The pair $(A, B)$ is controllable, the values of the elements of $A$ and $B$ are time-invariant and unknown. The plant sate vector is directly measurable, as well as the control signal $u$.

The reference model, which defines the required control quality for the closed-loop system with the control signal $u$ and the plant (2.1.1), takes the form:

$$\dot{x}_{ref} = A_{ref} x_{ref} + B_{ref} r, \quad x_{ref}(0) = x_{0ref}, \quad (2.1.2)$$

where $x_{ref} \in R^n$ is the reference model state vector, $x_{0ref} \in R^n$ is the vector of the initial conditions, $r \in R$ is the reference signal, $A_{ref} \in R^{n \times n}$ is a Hurwitz state matrix of the reference model, $B_{ref} \in R^n$ is the reference model input vector.

The control law for the plant (2.1.1) is chosen as:

$$u = \hat{k}_x x + \hat{k}_r r, \qquad (2.1.3)$$

where $\hat{k}_x^T \in R^n$ and $\hat{k}_r \in R$ are adjustable parameters, and $\hat{k}_r(0) \neq 0$.

The control law (2.1.3) is substituted into (2.1.1) to obtain the closed-loop description:

$$\dot{x} = \left(A + B\hat{k}_x\right)x + B\hat{k}_r r. \qquad (2.1.4)$$

The Erzberger's conditions [19] are assumed to be met for (2.1.4).

**Assumption 2.** *There exist ideal values of the control law parameters $k_x^T \in R^n$ and $k_r \in R$ such that the following equalities hold:*

$$A + Bk_x = A_{ref}; \; Bk_r = B_{ref}. \qquad (2.1.5)$$

Then the error equation, which is obtained by the subtraction of the reference model equation (2.1.2) from the plant one (2.1.4), is written as:

$$\dot{e}_{ref} = A_{ref} e_{ref} + B\left[\hat{k}_x x + \hat{k}_r r\right] - \left(A_{ref} - A\right)x - B_{ref} r = A_{ref} e_{ref} + B\left[\tilde{k}_x x + \tilde{k}_r r\right]. \qquad (2.1.6)$$

Here $e_{ref} = x - x_{ref}$, $\tilde{k}_x = \hat{k}_x - k_x$, $\tilde{k}_r = \hat{k}_r - k_r$. Let the following notation be introduced into (2.1.6):

$$\omega = \begin{bmatrix} x^T & r \end{bmatrix}^T, \; \tilde{\theta}^T = \begin{bmatrix} \tilde{k}_x & \tilde{k}_r \end{bmatrix} = \hat{\theta}^T - \theta^T, \qquad (2.1.7)$$

where $\omega \in R^{n+1}$ is the regressor, $\tilde{\theta} \in R^{n+1}$ is the error vector of the control law adjustable parameters.

Considering (2.1.7) and the initial conditions, the equation (2.1.6) is rewritten in the same form as (1.1):

$$\dot{e}_{ref} = A_{ref} e_{ref} + B\tilde{\theta}^T \omega, \; e_{ref}(0) = x_0 - x_{0ref}. \qquad (2.1.8)$$

Considering the state feedback control problem, the derivation of the adaptive law (2.2) consists of two main steps: 1) to obtain the linear regression equation $Y(t) = \Delta(t)\theta$ with the measurable regressor $\Delta \in R$ and function $Y \in R^{n+1}$; 2) to transform this equation in order to obtain the regression (2.1) with regressor $\Omega$, which is functionally dependent from $\Delta$ and has the necessary properties according to Theorem.

To obtain the regression equation $Y(t) = \Delta(t)\theta$, the equation of the plant (2.1.1) is written in the form of a linear regression equation:

$$\dot{x} = \theta_{AB}^T \Phi$$
$$\theta_{AB}^T = \begin{bmatrix} A & B \end{bmatrix}, \; \Phi = \begin{bmatrix} x^T & u \end{bmatrix}^T, \qquad (2.1.9)$$

where $\Phi \in R^{n+1}$ is a measurable regressor, $\theta_{AB}^T \in R^{n \times (n+1)}$ is the vector of the unknown parameters.

Despite the fact that $\dot{x}$ is not directly measurable, let the stable linear filters be introduced for all dynamic quantities from (2.1.9):

$$\dot{\bar{\mu}} = -l\bar{\mu} + \dot{x}, \; \bar{\mu}(0) = 0_n,$$
$$\dot{\bar{\Phi}} = -l\bar{\Phi} + \Phi, \; \bar{\Phi}(0) = 0_{n+1}, \qquad (2.1.10)$$

where $l > 0$ is the filters constant.

The regressor $\bar{\Phi}$ is defined as a solution of the second differential equation from (2.1.10), whereas, according to [20], the function $\bar{\mu}$ can be calculated in the following way even in case we do not know the value of $\dot{x}$:

$$\bar{\mu} = e^{-lt}\bar{\mu}(0) + x(t) - e^{-lt}x(0) - l\bar{x}(t) + le^{-lt}\bar{x}(0), \quad (2.1.11)$$

where $\bar{x}$ is the element of the vector $\bar{\Phi}$.

Taking into consideration the filtration (2.1.10) and the fact that the initial conditions for (2.1.11) are unknown, the equation (2.1.9) is rewritten as:

$$\bar{z} = x - l\bar{x} = \bar{\mu} + e^{-lt}x(0) = \theta_{AB}^{T}\bar{\Phi} + e^{-lt}x(0) = \bar{\theta}_{AB}^{T}\bar{\varphi}, \quad \forall t \geq 0,$$
$$\bar{\varphi} = \begin{bmatrix} \bar{\Phi}^{T} & e^{-lt} \end{bmatrix}^{T}, \bar{\theta}_{AB}^{T} = \begin{bmatrix} A & B & x(0) \end{bmatrix}, \quad (2.1.12)$$

where $\bar{z}$ is the measurable function, $\bar{\varphi} \in R^{n+2}$ is the measurable regressor, $\bar{\theta}_{AB}^{T} \in R^{n \times (n+2)}$ is the augmented vector of the unknown parameters.

Having applied the DREM procedure [21, 22], the equation (2.1.12) with the vector regressor $\bar{\varphi}$ is transformed into the one with the scalar regressor. Following [21,22], the $(n+2)$ minimal-phase dynamic filters are introduced:

$$(.)_{f_i(t)} := [H_i(.)](t), \; H_i(p) = \frac{\alpha_i^f}{p + \beta_i^f}, \; i \in \{1, 2, \cdots, n+2\}, \quad (2.1.13)$$

where $\alpha_i^f, \beta_i^f > 0$ and $\alpha_i^f \neq \alpha_j^f$ for all $i \neq j$.

The function $\bar{z}$ and regressor $\bar{\varphi}$ are passed through (2.1.13) to form the extended regression equation:

$$\bar{z}_f(t) = \bar{\varphi}_f(t)\bar{\theta}_{AB}$$
$$\bar{z}_f(t) := \begin{bmatrix} \bar{z}(t) \; \bar{z}_{f_1}(t) \; \ldots \; \bar{z}_{f_{n+2}}(t) \end{bmatrix}^{T}, \bar{\varphi}_f(t) := \begin{bmatrix} \bar{\varphi}(t) \; \bar{\varphi}_{f_1}(t) \; \ldots \; \bar{\varphi}_{f_{n+2}}(t) \end{bmatrix}^{T}, \quad (2.1.14)$$

where $\bar{z}_f \in R^{(n+3) \times n}$, $\bar{\varphi}_f \in R^{(n+3) \times (n+2)}$.

The equation (2.1.1) is left-multiplied by $\text{adj}\{\bar{\varphi}_f^T \bar{\varphi}_f\}\bar{\varphi}_f^T$. Then the known equality $\text{adj}\{\bar{\varphi}_f^T \bar{\varphi}_f\}\bar{\varphi}_f^T \bar{\varphi}_f = \det\{\bar{\varphi}_f^T \bar{\varphi}_f\}I$ is applied to the obtained result to form the equation with the scalar regressor:

$$z(t) = \varphi(t)\bar{\theta}_{AB}$$
$$z(t) = \text{adj}\{\bar{\varphi}_f^T \bar{\varphi}_f\}\bar{\varphi}_f^T \bar{z}_f(t), \; \varphi(t) = \det\{\bar{\varphi}_f^T \bar{\varphi}_f\}, \quad (2.1.15)$$

where $z \in R^{(n+2) \times n}$, $\varphi \in R$.

Considering the definitions of $\bar{\theta}_{AB}$ and $\varphi \in R$, we form the following regression equations from (2.1.15):

$$z_A = z^T H = \varphi A, \; z_B = z^T e_{n+1} = \varphi B, \quad (2.1.16)$$

where $z_A \in R^{n \times n}$, $z_B \in R^n$, $H = \begin{bmatrix} I_{n \times n} & 0_{n \times 2} \end{bmatrix}^T \in R^{(n+2) \times n}$, $e_{n+1} = \begin{bmatrix} 0_{1 \times n} & 1 & 0 \end{bmatrix}^T \in R^{n+2}$.

Now it is possible to transform the regressions (2.1.16) to the regression with respect to the parameters $\theta$ of the control law (2.1.3). For this, let the Erzberger's conditions (2.1.5) be multiplied by the regressor $\varphi \in R$, and the functions (2.1.16) be substituted into the obtained result:

$$\bar{Y} = \theta\bar{\Delta}$$
$$\bar{\Delta} = \varphi B^T = z_B^T, \; \bar{Y} = \varphi\begin{bmatrix} A_{ref} - A & B_{ref} \end{bmatrix}^T = \begin{bmatrix} \varphi A_{ref} - z_A & \varphi B_{ref} \end{bmatrix}^T, \quad (2.1.17)$$

where $\bar{Y} \in R^{(n+1) \times n}$, $\bar{\Delta} \in R^{1 \times n}$.

The equation (2.1.17) is multiplied by $\bar{\Delta}^T$ to obtain the scalar regressor $\Delta$ from the

vector one $\bar{\Delta}$:

$$Y(t) = \Delta(t)\theta$$
$$Y = \bar{Y}(t)\bar{\Delta}^{\mathrm{T}}(t),\ \Delta(t) = \bar{\Delta}\bar{\Delta}^{\mathrm{T}},$$ (2.1.18)

where $Y \in R^{n+1}$, $\Delta \in R$.

**Remark 2.** *According to the results of Lemma 6.8 [2], as the filters* (2.1.10) *are stable, if* $\Phi \in \mathrm{FE}$, *then* $\bar{\varphi} \in \mathrm{FE}$ *as well. In* [22] *the implication* $\bar{\varphi} \in \mathrm{FE} \Rightarrow \varphi \in \mathrm{FE}$ *is proved for the extension scheme* (2.1.13)-(2.1.15). *Then, as the regressor* $\bar{\Delta}$ *depends from only one dynamic quantity* $\varphi$, *and, following Assumption* 1, *it does not become singular when* $\varphi = const$, *then it also holds that* $\varphi \in \mathrm{FE} \Rightarrow \bar{\Delta} \in \mathrm{FE} \Rightarrow \Delta \in \mathrm{FE}$. *As a result, when the procedure* (2.1.9)-(2.1.18) *is applied, the initial regressor* $\Phi$ *excitation does not vanish, and it holds that* $\Phi \in \mathrm{FE} \Rightarrow \Delta \in \mathrm{FE}$.

The next aim is to transform the equation (2.1.18) to obtain the regression (2.1) with the regressor $\Omega$, which meets all the requirements of Theorem.

Taking into consideration the results of [20, 23, 24], the filter with exponential forgetting is introduced:

$$\begin{cases} \dot{\beta} = \sigma,\ \beta(0) = 0 \\ \dot{v}_f = \exp(-\beta)v,\ v_f(0) = 0 \end{cases}$$ (2.1.19)

where $\sigma > 0$ is the arbitrary parameter, $v$ and $v_f$ are the input and output of the filter respectively.

The extended regressor $\Delta^2$ and function $\Delta Y$ are passed through (2.1.19):

$$\Upsilon = \Omega\theta$$
$$\Omega(t) = \int_{t_r^+}^{t} e^{-\sigma\tau}\Delta^2(\tau)d\tau,\ \Upsilon(t) = \int_{t_r^+}^{t} e^{-\sigma\tau}\Delta(\tau)Y(\tau)d\tau,$$ (2.1.20)

where $\Upsilon \in R^{n+1}$, $\Omega \in R$.

The following proposition with respect to the regressor $\Omega$ has been proved in [23]:

**Proposition 1.** *If* $\Delta \in \mathrm{FE}$ *over the time range* $\left[t_r^+;t_e\right]$ *and* $\Delta \in L_\infty$, *then*

1) $\forall t \geq t_r^+\ \Omega(t) \in L_\infty,\ \Omega(t) \geq 0;$

2) $\forall t \geq t_e\ \Omega(t) > 0,\ \int_{t_r^+}^{t_e} e^{-\sigma\tau}\Delta^2(\tau)d\tau \leq \Omega(t) \leq \delta_\Delta^2\sigma^{-1}.$

where $\delta_\Delta = \sup_{t \geq 0}\max|\Delta(t)|$.

*The proof of Proposition* 1 *could be found in* [23, 24].

Proposition 1 is to consider the case when $\Delta \in L_\infty$, but, from the point of view of the adaptive control systems practical application, it is impossible to claim a priori that $\Delta \in L_\infty$, since $\Delta$ depends from the state vector $x$ (see (2.1.9)-(2.1.18)), Therefore, in the next proposition we consider the situation when $\Delta \notin L_\infty$, but it does not grow faster than some known exponent.

**Proposition 2.** *If* $\Delta \in \mathrm{FE}$ *over the time range* $\left[t_r^+;t_e\right]$, *and* $\forall t \geq t_r^+\ |\Delta| \leq c_1 e^{c_2 t}$, *and* $\sigma > 2c_2$, *then*:

1) $\forall t \geq t_r^+\ \Omega(t) \in L_\infty,\ \Omega(t) \geq 0;$

2) $\forall t \geq t_e \ \Omega(t) > 0, \ \dfrac{c_1^2}{c_3}\left(e^{-c_3 t_r^+} - e^{-c_3 t_e}\right) \leq \Omega(t) \leq \dfrac{c_1^2}{c_3}.$

where $c_1 > 0$, $c_2 > 0$, $c_3 = \sigma - 2c_2 > 0$.

*The Proof of Proposition 2 is postponed to Appendix.*

According to Proposition 2, it is always possible to choose a parameter $\sigma$ value such that, even if $\Delta \notin L_\infty$, the regressor $\Omega \in L_\infty$ and satisfies the requirements of Theorem. In practice, it is possible to set a majorant function in the form of the conservative exponent $|\Delta| \leq c_1 e^{c_2 t}$ with the known $c_1$ and $c_2$ for the regressor $\Delta$ of any unstable plant, and therefore the requirements of Proposition 2 are not restrictive.

The results of the second clauses of Proposition 1 and 2 are summarized by the introduction of an inequality for $\Omega$, which holds $\forall t \geq t_e$ and for any $\Delta \in$ FE such that $\Delta \in L_\infty$ or $|\Delta| \leq c_1 e^{c_2 t}$:

$$0 < \underbrace{\min\left\{\int_{t_r^+}^{t_e} e^{-\sigma\tau}\Delta^2(\tau)d\tau;\ \dfrac{c_1^2}{c_3}\left(e^{-c_3 t_r^+} - e^{-c_3 t_e}\right)\right\}}_{\Omega_{LB}} \leq \Omega(t) \leq \underbrace{\max\left\{\delta_\Delta^2 \sigma^{-1};\ \dfrac{c_1^2}{c_3}\right\}}_{\Omega_{UB}}. \quad (2.1.21)$$

According to the equation (2.1.21), if the condition $\Delta \in$ FE is met, the regressor $\Omega$ has the necessary properties from the point of view of Theorem. Hence, if the parameters of the control law (2.1.3) are adjusted according to (2.2), then the error $\xi$ is exponentially stable, as far as the state feedback adaptive control problems are concerned.

Thus, the state feedback adaptive control system with the developed adaptive law (2.2) consists of the control law (2.1.3) as well as the procedures (2.1.9)-(2.1.18) to obtain the linear regression $Y(t) = \Delta(t)\theta$ and transform it (2.1.19)-(2.1.20).

Compared to the adaptive law (1.3), the proposed one (2.2) provides exponential convergence of the error $\xi$ to zero under the weaker condition $\Delta \in$ FE. Compared to the various composite adaptive laws [5, 8-10], which have become widespread in recent years, the proposed law (2.2) provides elementwise monotonicity of the controller adjustable parameters $\hat{\theta}$. It does not require to know the sign or the values of the elements of the vector $B$. It is less sensitive to the value of the manually chosen adaptive gain $\gamma$. The obtained result can be directly generalized to the case of MIMO LTI plants.

## 2.2. Output feedback control case

Let a classical problem of output-feedback adaptive control of LTI SISO plants be considered [1-3]:

$$y(t) = b_m \dfrac{Z(p)}{R(p)} u(t), \quad (2.2.1)$$

where $p = d/dt$ is the differentiation operator, $y$ is the plant output, $u$ is the control signal, $b_m$ is the plant gain, $Z(p) = p^m + \sum_{i=0}^{m-1} b_i b_m^{-1} p^i$ and $R(p) = p^n + \sum_{i=0}^{n-1} a_i p^i$ are the characteristic polynomials with quasi-stationary $\left(\dot{b}_i \approx 0,\ \dot{a}_i \approx 0\right)$ unknown parameters.

The required control quality for the plant (2.2.1) is defined as the following reference model:
$$y_{ref}(t) = b_{ref} \frac{Z_{ref}(p)}{R_{ref}(p)} r(t), \quad (2.2.2)$$

where $y_{ref}$ is the output of the reference model, $r$ is the reference signal, $b_{ref}$ is the reference model gain, $Z_{ref}(p)$ and $R_{ref}(p)$ are Hurwitz characteristic polynomials, which degrees are $m^*$ and $n^*$ respectively. The relative degree of the reference model $\rho^* = n^* - m^*$ is assumed to be equal to the relative degree of the plant $\rho = n - m$.

The control aim for the plant (2.2.1) is to ensure that the plant output (2.2.1) tracks asymptotically the reference model output (2.2.2):
$$\lim_{t \to \infty}(y(t) - y_{ref}(t)) = \lim_{t \to \infty} \varepsilon(t) = 0. \quad (2.2.3)$$

This problem is considered under the following classical assumptions [1-3, 13, 14].

**Assumption 2.** *The numerator polynomial $Z(p)$ is a Hurwitz one.*

**Assumption 3.** *The values of n and m are known. So, the plant relative degree $\rho = n - m \geq 1$ is also known.*

**Assumption 4.** *Only the signals y and u are directly measurable, but their derivatives are not.*

To parameterize the adaptive control problem and eventually obtain the error equation, the state filters are introduced for the plant (2.2.1) according to [1-3]:
$$\begin{aligned} \dot{v}_1 &= \Lambda v_1 + hu, \quad v_1(0) = 0, \\ \dot{v}_2 &= \Lambda v_2 + hy, \quad v_2(0) = 0, \end{aligned} \quad (2.2.4)$$

where $v_1 \in R^{n-1}$, $v_2 \in R^{n-1}$, $h = [0,0,...,0,1]^T \in R^{n-1}$, $\Lambda$ is the companion matrix of the Hurwitz polynomial $\Lambda(p) = \Lambda_0(p) Z_{ref}(p)$.

Then, considering the filters (2.2.3), the plant transfer function (2.2.1) can be transformed [1-3] into:
$$y = \frac{Z_{ref}(p)}{R_{ref}(p)} b_m \left[u - k_1^T v_1 - k_2^T v_2 - k_3 y\right] + \varepsilon_y \quad (2.2.5)$$

where $k_1 \in R^{n-1}$, $k_2 \in R^{n-1}$, $k_3 \in R$, $\varepsilon_y$ is an exponentially vanishing disturbance, which is caused by the fact that the initial conditions of the plant and filters do not coincide to each other.

Using, (2.2.5), the following control law with the adjustable parameters is chosen to achieve the goal (2.2.3):
$$u = \hat{k}_1^T v_1 + \hat{k}_2^T v_2 + \hat{k}_3 y + \hat{k}_4 r, \quad (2.2.6)$$

where $\hat{k}_4(0) \neq 0$.

Considering (2.2.6), the reference model equation (2.2.2) is subtracted from the plant one (2.2.5):
$$\begin{aligned} \varepsilon &= \frac{Z_{ref}(p)}{R_{ref}(p)} b_m \left[\left(\hat{k}_4 - b_m^{-1} b_{ref}\right) r + \tilde{k}_1^T v_1 + \tilde{k}_2^T v_2 + \tilde{k}_3 y\right] + \varepsilon_y = \\ &= \frac{Z_{ref}(p)}{R_{ref}(p)} b_m \left[\tilde{k}_4 r + \tilde{k}_1^T v_1 + \tilde{k}_2^T v_2 + \tilde{k}_3 y\right] + \varepsilon_y, \end{aligned} \quad (2.2.7)$$

where $k_4 = b_m^{-1} b_{ref}$.

The transfer function description of (2.2.7) is transformed into the state-space one to obtain the differential error equation with the measurable output (1.2):

$$\dot{e}_{ref} = A_{ref} e_{ref} + B \tilde{\theta}^T \omega, \ e_{ref}(0) = e_{0ref}, \qquad (2.2.8)$$
$$\varepsilon = c^T e_{ref},$$

where $A_{ref} \in R^{n^* \times n^*}$ is the companion matrix of the polynomial $R_{ref}(p)$, $B \in R^{n^*}$ is the gain vector, $c \in R^{n^*}$ is the output vector, $\omega = [r \ v_1 \ v_2 \ y]^T \in R^{2n}$ is the regressor vector, $\tilde{\theta} = [\tilde{k}_4 \ \tilde{k}_1^T \ \tilde{k}_2^T \ \tilde{k}_3]^T \in R^{2n}$ is the vector of the parametric errors.

Considering the problem of output feedback adaptive control, the derivation of the adaptive law (2.2) also includes two steps: 1) to obtain the linear regression equation $Y(t) = \Delta(t)\theta$ with the measurable regressor $\Delta \in R$ and function $Y \in R^{2n}$; 2) to transform this equation in order to obtain the regression (2.1) with the regressor $\Omega$, which is functionally dependent from $\Delta$ and has the required properties according to Theorem.

To obtain a linear regression equation $Y(t) = \Delta(t)\theta$, a control law with ideal parameters is introduced:

$$u^* = k_1^T v_1 + k_2^T v_2 + k_3 y + k_4 r = k_1^T \frac{\alpha(p)}{\Lambda(p)} u + k_2^T \frac{\alpha(p)}{\Lambda(p)} y + k_3 y + k_4 r, \qquad (2.2.9)$$

where $\alpha(p)$ is the differentiation operator, which is defined as follows:

$$\alpha(p) = \begin{cases} [p^{n-2}, p^{n-3}, \ldots, p, 1]^T & \text{if } n \geq 2, \\ 0 & \text{if } n = 1. \end{cases} \qquad (2.2.10)$$

Substituting (2.2.9) into (2.2.1), the equation of the closed-loop is obtained:

$$y(t) = \frac{k_4 b_m Z(p) \Lambda(p)}{[\Lambda(p) - k_1^T \alpha(p)] R(p) - b_m Z(p) [k_2^T \alpha(p) + k_3 \Lambda(p)]} r(t). \qquad (2.2.11)$$

Considering $\Lambda(p) = \Lambda_0(p) Z_{ref}(p)$, the equations (2.2.11) and (2.2.2) are set equal to obtain a kind of the Erzberger's conditions (2.1.5):

$$b_{ref} k_1^T \alpha(p) R(p) + b_{ref} b_m Z(p) [k_2^T \alpha(p) + k_3 \Lambda(p)] + \\ + k_4 b_m Z(p) \Lambda_0(p) R_{ref}(p) = b_{ref} \Lambda(p) R(p). \qquad (2.2.12)$$

**Remark 3.** *According to the results of [1-2, 25, 26], under the conditions of Assumption 3 the equation (2.2.12) is solvable with respect to the parameters $k_1, k_2, k_3, k_4$, and such solution is unique.*

Then, to obtain the equation $Y(t) = \Delta(t)\theta$ on the basis of (2.2.12), the plant (2.2.1) is transformed into the observability canonical form:

$$\begin{cases} \dot{x} = A_o x + B_o u, \ x(0) = x_0 \\ y = C^T x, \end{cases};$$

$$A_o = \begin{bmatrix} -a & I_{(n-1)\times(n-1)} \\ & 0_{1\times(n-1)} \end{bmatrix}; \ B_o = \begin{bmatrix} 0_{n-(m+1)} \\ b \end{bmatrix}; \ C^T = \begin{bmatrix} 1 & 0_{1\times(n-1)} \end{bmatrix}, \qquad (2.2.13)$$

where $x \in R^n$ is directly unmeasurable state vector, $a \in R^n = [a_{n-1} \ldots a_0]^T$, $b \in R^{m+1} = [b_m \ldots b_0]^T$.

Let a Hurwitz matrix of canonical form $\Psi_c \in R^{n \times n}$ be introduced. Considering the equality $(\psi - a)C^T = A_o - \Psi_c$, the expression $\Psi_c x$ is added to and subtracted ($\pm$) from (2.2.13) to obtain:

$$\dot{x} = \Psi_c x + (\psi - a)y + B_o u. \tag{2.2.14}$$

where $\psi \in R^n$ is the coefficients vector of the characteristic polynomial of the matrix $\Psi_c$.

To form the regression with respect to the parameters $a$ and $b$ from (2.2.13), the Kreisselmeier's parametrization [27] is applied. By analogy with (2.2.4), the set of the state linear filters is introduced:

$$\begin{aligned}\dot{\eta}_{f1} &= \Psi_c^T \eta_{f1} + Cu, \ \eta_{f1}(0) = 0_n, \\ \dot{\eta}_{f2} &= \Psi_c^T \eta_{f2} + Cy, \ \eta_{f2}(0) = 0_n,\end{aligned} \tag{2.2.15}$$

where $\eta_{f1}, \eta_{f2} \in R^n$.

Then, according to the results of [27], the regression equation is obtained:

$$\begin{aligned}x &= \Phi \theta_{(\psi-a)B_o} + e^{\Psi_c t} x(0), \\ \Phi &= [\Phi_{f_1} \ldots \Phi_{f_{2n}}], \ \Phi_{f_i} = T_i \eta_{f1}, \ \Phi_{f_{i+n}} = T_i \eta_{f2}, \ i = \overline{1,n}.\end{aligned} \tag{2.2.16}$$

where $\Phi \in R^{n \times 2n}$ is the measurable regressor, $\theta_{(\psi-a)B_o} \in R^{2n} = [\psi^T - a^T \ B_o^T]^T$ is the vector of the unknown parameters, $T_i \in R^{n \times n}$ is the transformation matrix, which is formed from the coefficients of the numerator polynomial of the vector function $(sI - \Psi_c)^{-1} e_i$, $e_i$ is a vector, which is filled with zeros, but its $i^{th}$ element is equaled to one.

The equation (2.2.16) is multiplied by $C^T$ to obtain the measurable regression function:

$$\begin{aligned}\bar{z} &= y - C^T \Phi \theta_\psi = C^T \Phi \theta_{(\psi-a)B_o} - C^T \Phi \theta_\psi + C^T e^{\Psi_c t} x(0) = \bar{\theta}_{-aB_o}^T \bar{\varphi}, \\ \bar{\varphi} &= [C^T \Phi \ \ C^T e^{\Psi_c t}]^T, \ \bar{\theta}_{-aB_o} = [-a^T \ B_o^T \ x(0)]^T \ \theta_\psi = [\psi^T \ 0_{1 \times n}]^T.\end{aligned} \tag{2.2.17}$$

where $\bar{\varphi} \in R^{3n}$, $\bar{\theta}_{-aB_o} \in R^{3n}$.

To obtain a regression with the scalar regressor $\Delta$ from (2.2.17), the procedure from [22, 28] is applied. In accordance with it, the following extension scheme is introduced:

$$\begin{aligned}\dot{\bar{z}}_f^T &= -l \cdot \bar{z}_f^T + \bar{z} \bar{\varphi}^T, \ \bar{z}_f(0) = 0, \\ \dot{\bar{\varphi}}_f^T &= -l \cdot \bar{\varphi}_f^T + \bar{\varphi} \bar{\varphi}^T, \ \bar{z}_f(0) = 0,\end{aligned} \tag{2.2.18}$$

where $l > 0$.

The function $\bar{z}$ and regressor $\bar{\varphi}$ are passed through (2.2.18) to obtain the extended regression equation:

$$\bar{z}_f(t) = \bar{\varphi}_f(t) \bar{\theta}_{-aB_o}, \tag{2.2.19}$$

where $\bar{z}_f \in R^{3n}$, $\bar{\varphi}_f \in R^{3n \times 3n}$.

The equation (2.2.19) is left-multiplied by $\text{adj}\{\bar{\varphi}_f\}$ and, using the equality $\text{adj}\{\bar{\varphi}_f\} \bar{\varphi}_f = \det\{\bar{\varphi}_f\} I$, it is obtained:

$$z(t) = \varphi(t)\overline{\theta}_{-aB_o}$$
$$z(t) = \mathrm{adj}\{\overline{\varphi}_f\}\overline{z}_f(t),\ \varphi(t) = \det\{\overline{\varphi}_f\},$$
(2.2.20)

where $\varphi,\ z \in R^{3n}$ are the measurable regressor and function.

**Remark 4.** *It is also possible to use the scheme (2.1.13)-(2.1.14) to obtain the extended regression (2.2.19), which adds a degree of freedom to the algorithm (2.2.20). The advantage of the scheme (2.2.18)-(2.2.19) as compared to (2.1.13)-(2.1.14) is that lower number of parameters need to be chosen. The extension scheme (2.2.18)-(2.2.19) can also be used instead of (2.1.13)-(2.1.14) for the state feedback control problem.*

Given the definition $\overline{\theta}_{-aB_0}$ and $\varphi \in R$ from (2.2.20), it is easy to obtain the regression equations:
$$z_a = H_1 z = \varphi a,\ z_b = H_2 z = \varphi b,$$
$$H_1 = \begin{bmatrix} -I_{n\times n} & 0_{n\times 2n} \end{bmatrix},\ H_2 = \begin{bmatrix} 0_{(m+1)\times(2n-(m+1))} & I_{(m+1)\times(m+1)} & 0_{(m+1)\times n} \end{bmatrix}$$
(2.2.21)

where $z_a \in R^n,\ z_b \in R^{m+1}$ are measurable functions, $H_1 \in R^{n\times 3n},\ H_2 \in R^{(m+1)\times 3n}$ are transformation matrices.

Then, the condition (2.2.12) are multiplied by φ. The coefficients in the left and right hand sides of the resulting equation with the corresponding degrees of the differentiation operator *p* are considered to be equal. The corresponding scalar equations from (2.2.21) with respect to $a_{n-1}\ldots a_0$ and $b_m\ldots b_0$ are substituted into the obtained result. Then a matrix regression equation with respect to the unknown parameters θ of the ideal control law (2.2.9) is written:
$$N = M\theta,$$
(2.2.22)

where $M \in R^{2n\times 2n},\ N \in R^{2n}$ are the measurable regressor and function.

Then, the matrix regressor *M* is to be transformed into the scalar one $\Delta$. For that the equation (2.2.22) is multiplied by the matrix $\mathrm{adj}\{M\}$. After that the equality $\mathrm{adj}\{M\}M = \det\{M\}I$ is applied to obtain:
$$Y(t) = \Delta(t)\theta$$
$$Y(t) = \mathrm{adj}\{M(t)\}N(t),\ \Delta(t) = \det\{M(t)\},$$
(2.2.23)

where $Y \in R^{2n},\ \Delta \in R$ is the scalar regressor.

**Remark 5.** *According to the results of [28], when the extension scheme (2.2.18)-(2.2.20) is applied, the implication $\overline{\varphi} \in \mathrm{FE} \Rightarrow \varphi \in \mathrm{FE}$ holds. The regressor M is defined by a single dynamic quantity $\varphi$. In accordance with Remark 3, when $\varphi = const$, M does not become singular. So, $\varphi \in \mathrm{FE} \Rightarrow M \in \mathrm{FE} \Rightarrow \Delta \in \mathrm{FE}$ holds. Thus, considering the parameterization (2.2.13)-(2.2.23), the excitation of the original regressor $\overline{\varphi}$ does not vanish and $\overline{\varphi} \in \mathrm{FE} \Rightarrow \Delta \in \mathrm{FE}$ holds.*

The regression (2.2.23) is to be transformed into (2.2.1) with the regressor $\Omega$, which has the required properties defined by Theorem. Using the results of [20, 23, 24], the filter with the exponential forgetting (2.1.19) is introduced.

The extended regressor $\Delta^2$ and function $\Delta Y$ are passed through the filter (2.1.19) to obtain:

$$\Upsilon = \Omega\theta$$

$$\Omega(t) = \int_{t_r^+}^{t} e^{-\sigma\tau}\Delta^2(\tau)d\tau, \ \Upsilon(t) = \int_{t_r^+}^{t} e^{-\sigma\tau}\Delta(\tau)Y(\tau)d\tau, \tag{2.2.24}$$

where $\Upsilon \in R^{2n}$, $\Omega \in R$.

Using the proved clauses of Proposition 1 and 2, when $\Delta \in$ FE, the regressor $\Omega$ has all properties required by Theorem as it can be seen from the following proposition.

**Proposition 3.** *If* $\Delta \in$ FE *over the time interval* $\left[t_r^+; t_e\right]$ *and* $\Delta \in L_\infty$ *or* $|\Delta| \leq c_1 e^{c_2 t}$ *and* $\sigma > 2c_2$, *then*

$$1) \ \forall t \geq t_r^+ \ \Omega(t) \in L_\infty, \Omega(t) \geq 0;$$
$$2) \ \forall t \geq t_e \ \Omega(t) > 0, \ 0 < \Omega_{LB} \leq \Omega(t) \leq \Omega_{UB}.$$

*where* $\delta_\Delta = \sup_{t \geq 0} \max |\Delta(t)|$, $c_3 = \sigma - 2c_2 > 0$.

The correctness of Proposition 3 follows from the proofs of Propositions 1 and 2, and the estimates of $\Omega_{LB}$ and $\Omega_{UB}$, which coincide with the ones given in (2.1.21).

Then, if the parameters of the control law (2.2.6) are adjusted according to (2.2), the error $\xi$ is exponentially stable, as far as the output feedback control problem is concerned.

Thus, the output feedback adaptive control system with the developed adaptive law (2.2) consists of the control law (2.2.6) and the procedures to obtain (2.2.13)-(2.2.23) the linear regression $Y(t) = \Delta(t)\theta$ and transform it (2.2.24).

Compared to the law (1.5), the proposed one (2.2) does not require the transfer function $c^T(pI - A_{ref})^{-1} B$ to be SPR. It also provides exponential convergence of the error $\xi$ to zero under the weaker condition $\Delta \in$ FE. Compared to other approaches [3, 13, 14], in addition to overcoming the SPR requirement, the proposed law (2.2): 1) provides elementwise monotonicity of the adjustable parameters $\hat{\theta}$ of the control law, 2) is substantially easier to be implemented in practice, 3) does not require to know the sign or value of the elements of the plant gain $b_m$, and 4) is less sensitive to the manually chosen adaptive gain $\gamma$ value.

### 3. Numerical Experiments

The numerical experiments with the state and output feedback adaptive control systems proposed in the second section of the paper have been conducted in Matlab/Simulink. The numerical integration by the Euler method with a constant step size of $\tau_s = 10^{-4}$ seconds was used.

### 3.1 State feedback adaptive control system

An unstable aperiodic link of second order was chosen as a plant to conduct experiments to validate effectiveness of the proposed state feedback adaptive control system:

$$\dot{x} = \begin{bmatrix} 0 & 1 \\ 4 & 2 \end{bmatrix} x + \begin{bmatrix} 0 \\ 2 \end{bmatrix} u, \ x(0) = \begin{bmatrix} 0 \\ 0 \end{bmatrix}. \tag{3.1.1}$$

The required control quality was defined as the following reference model:

$$\dot{x}_{ref} = \begin{bmatrix} 0 & 1 \\ -8 & -4 \end{bmatrix} x_{ref} + \begin{bmatrix} 0 \\ 8 \end{bmatrix} r, \ x_{ref}(0) = \begin{bmatrix} 0 \\ 0 \end{bmatrix}. \tag{3.1.2}$$

The reference $r$ value, the filters (2.1.10), (2.1.13), (2.1.18) parameters, the values of $\gamma_0$ and $\gamma_1$, the initial values of the control law (2.1.3) parameters were chosen as:

$$r = 1;\ l = 1;\ \alpha_i^f = \beta_i^f = i;\ i \in \overline{1,5};\ \sigma = \tfrac{5}{10};\ \gamma_0 = 1;\ \gamma_1 = 0;\ \hat{\theta}(0) = \begin{bmatrix} 0 & 0 & 1 \end{bmatrix}^T. \quad (3.1.3)$$

Figure 1 presents the transients of: (a) the elements of the plant (3.1.1) and the reference model (3.1.2) state vectors, (b) the adjustable parameters of the control law, (c) the regressor $\Omega$, and (d) the value of $\lambda_{max}(\omega\omega^T)$.

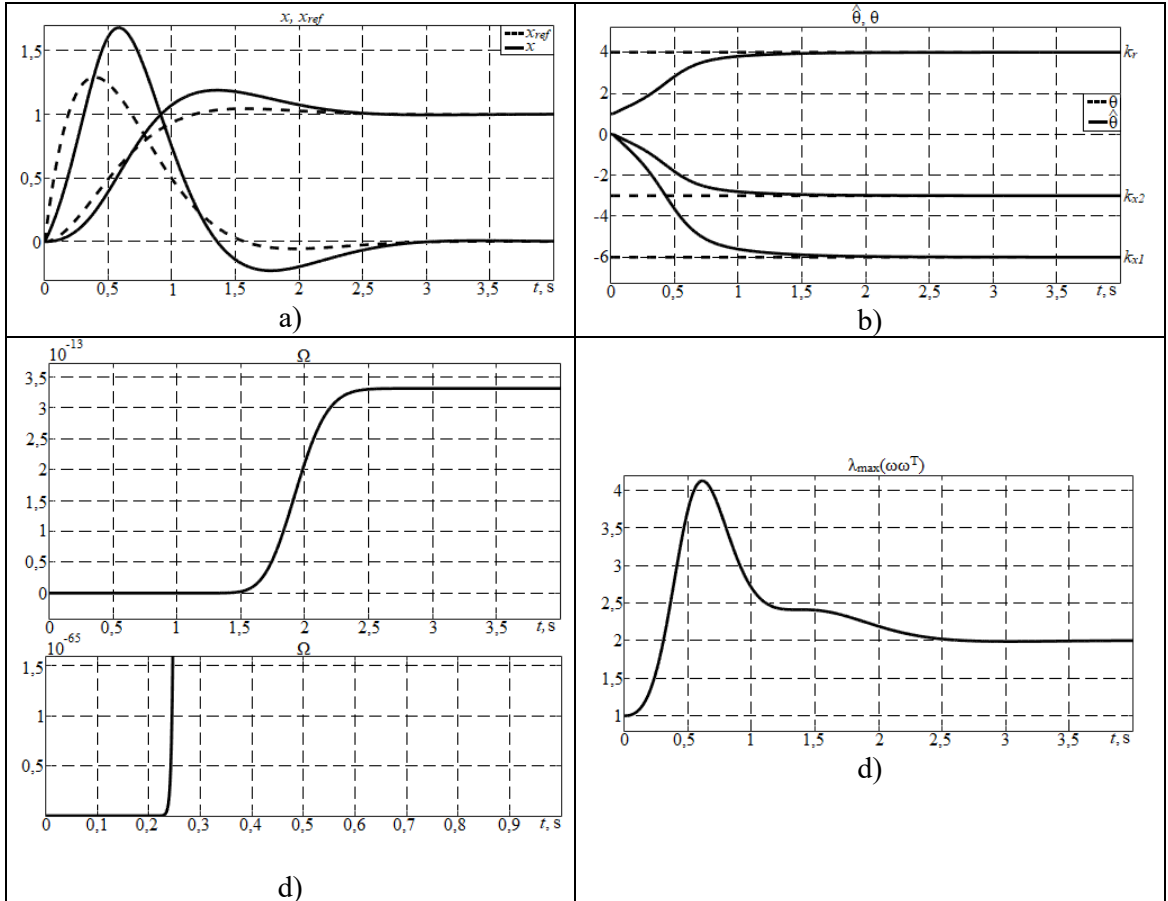

Figure 1. Transients of: (a) elements of plant (3.1.1) and reference model (3.1.2) state vectors, (b) adjustable parameters of control law, (c) regressor $\Omega$, and (d) value of $\lambda_{max}(\omega\omega^T)$

The simulation results confirmed the theoretical conclusions and verified the exponential stability of the augmented tracking error $\xi$ when the proposed tuning law (2.2) was applied.

Then, it was demonstrated that the adaptive law (2.2) was invariant to the sign of the vector $B$. For this purpose, the initial values of the control law parameters were set as follows: $\hat{\theta}(0) = \begin{bmatrix} 0 & 0 & -1 \end{bmatrix}^T$. Figure 2 shows (a) the transients of the plant (3.1.1) and the reference model (3.1.2) state vector elements, and (b) the adjustable parameters $\hat{\theta}$ of the control law.

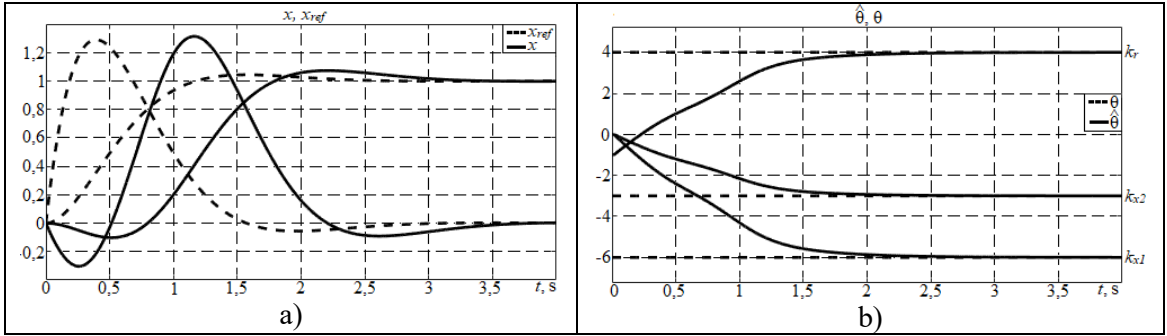

Figure 2. Transients of (a) plant (3.1.1) and reference model (3.1.2) state vector elements, and (b) adjustable parameters of control law

The transients shown in Fig. 2 confirmed the correctness and efficiency of the adaptive law (2.2) under the condition that the sign of the gain vector $B$ was unknown.

Then the developed system was simulated in the practically important stabilization mode (Figure 3). For this purpose, the plant initial conditions were set as $x(0) = \begin{bmatrix} 1 & 0 \end{bmatrix}^T$ and the reference signal was selected as $r = e^{-1t}$. The initial conditions of the reference model and other loop parameters were set according to (3.1.2) and (3.1.3).

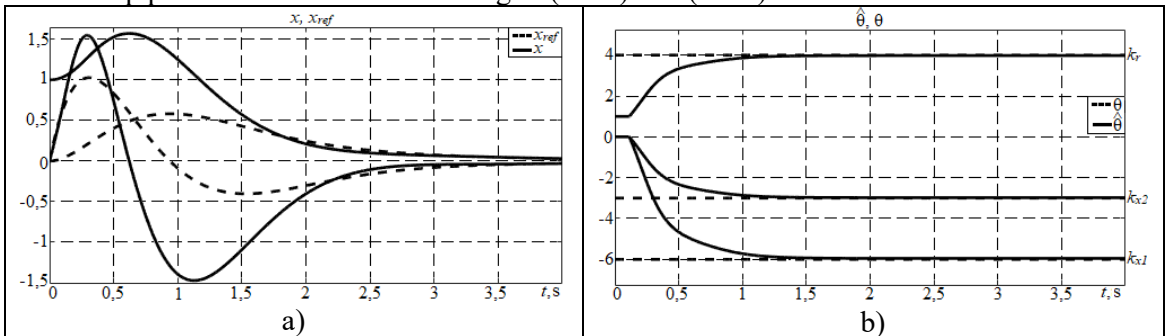

Figure 3. Transients of (a) plant (3.1.1) and reference model (3.1.2) state vector elements, and (b) adjustable parameters of control law

The results of the experiments, which are presented in Figure 3, demonstrated the effectiveness of the proposed adaptive law (2.2) to solve the problem of state feedback adaptive stabilization.

The next experiment was to compare the performance of the proposed system for different values of $\gamma_0$. Fig. 4 shows the error norms $\|\varepsilon\|$, which were obtained for different values of the parameter $\gamma_0$.

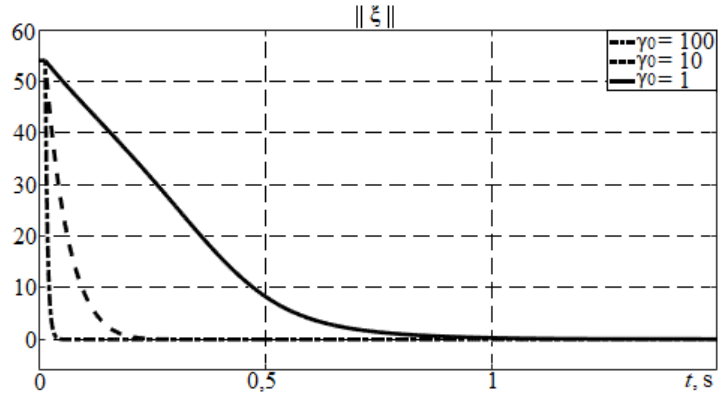
Figure 4. The dependence of $\|\xi\|$ from value of parameter $\gamma_0$

The transients shown in Fig. 4 confirmed the conclusions made in the proof of Theorem. They also demonstrated that the rate of the error $\xi$ convergence to zero could be improved if the value of $\gamma_0$ was made higher.

Finally, it was checked whether the developed adaptive control system allowed one to set the value of the minimal rate of the error $\xi$ convergence to zero with the help of choice of $\gamma_1$ value, as it was noted in Remark 1. For this purpose, the developed system was modelled using the values $\gamma_1 = 0$, $\gamma_1 = 10$ and different reference signals $r$. Figure 5 presents (a) transients of the error norm $\|\xi\|$ when $\gamma_1 = 0$ and different $r$ were used, (b) $\gamma_1 = 10$ and the same set of $r$ were used.

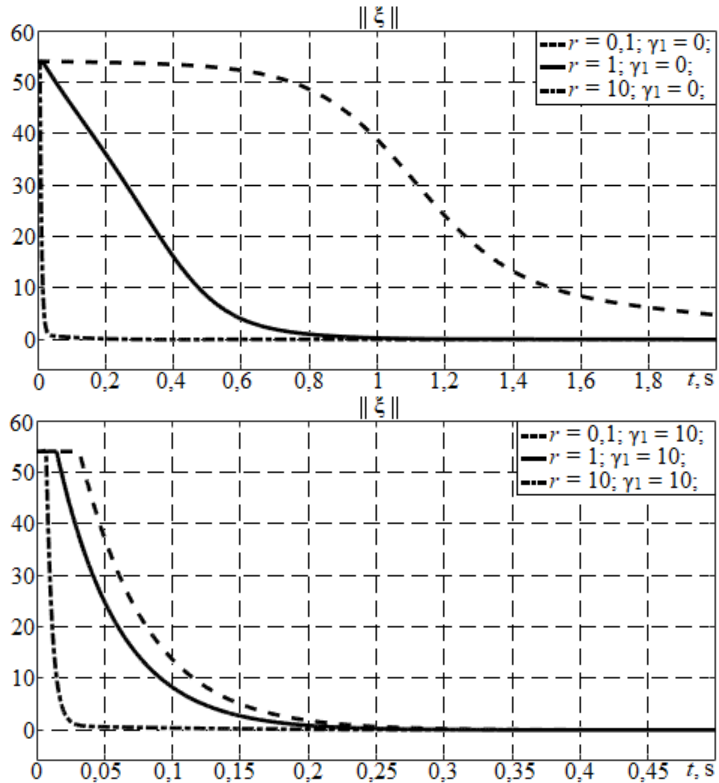
Figure 5. Dependence of $\|\xi\|$ from $r$ and $\gamma_1$

The results of the experiment shown in Fig. 5 confirmed that the value of the minimal rate of the error $\|\xi\|$ convergence to zero could be adjusted by change of the $\gamma_1$ value.

### 3.2 Output feedback adaptive control system

The plant and reference model to solve the output feedback adaptive control problem were chosen as (3.1.1) и (3.2.2), but they were written as transfer functions:

$$y = \frac{2}{p^2 - 2p - 4}u; \quad y_{ref} = \frac{8}{p^2 + 4p + 8}r. \tag{3.2.1}$$

The only directly measurable signals were the output $y = x_1$ and the control signal $u$.

The reference value $r$, the filters (2.2.4), (2.2.15), (2.2.16), (2.2.18), (2.2.24) parameters, the values of $\gamma_0$ and $\gamma_1$, the initial conditions for the outputs of the plant and reference model, and the initial values of control law (2.1.3) parameters were chosen as:

$$r = 1; \Lambda = -1; \psi = \begin{bmatrix} 20 & 100 \end{bmatrix}^T; l = 0,1;$$

$$T_1 = \begin{bmatrix} 1 & 0 \\ 0 & -100 \end{bmatrix}; T_2 = \begin{bmatrix} 0 & 1 \\ 1 & 20 \end{bmatrix}; y(0) = y_{ref}(0) = 0; \tag{3.2.2}$$

$$\sigma = \tfrac{5}{10}; \gamma_0 = 1; \gamma_1 = 0; \hat{\theta}(0) = \begin{bmatrix} 1 & 0 & 0 & 0 \end{bmatrix}^T.$$

Figure 6 depicts the transients of (a) the plant $y$ and the reference model $y_{ref}$ outputs, (b) the adjustable parameters $\hat{\theta}$, (c) the regressor $\Omega$, and (d) the value of $\lambda_{max}(\omega\omega^T)$.

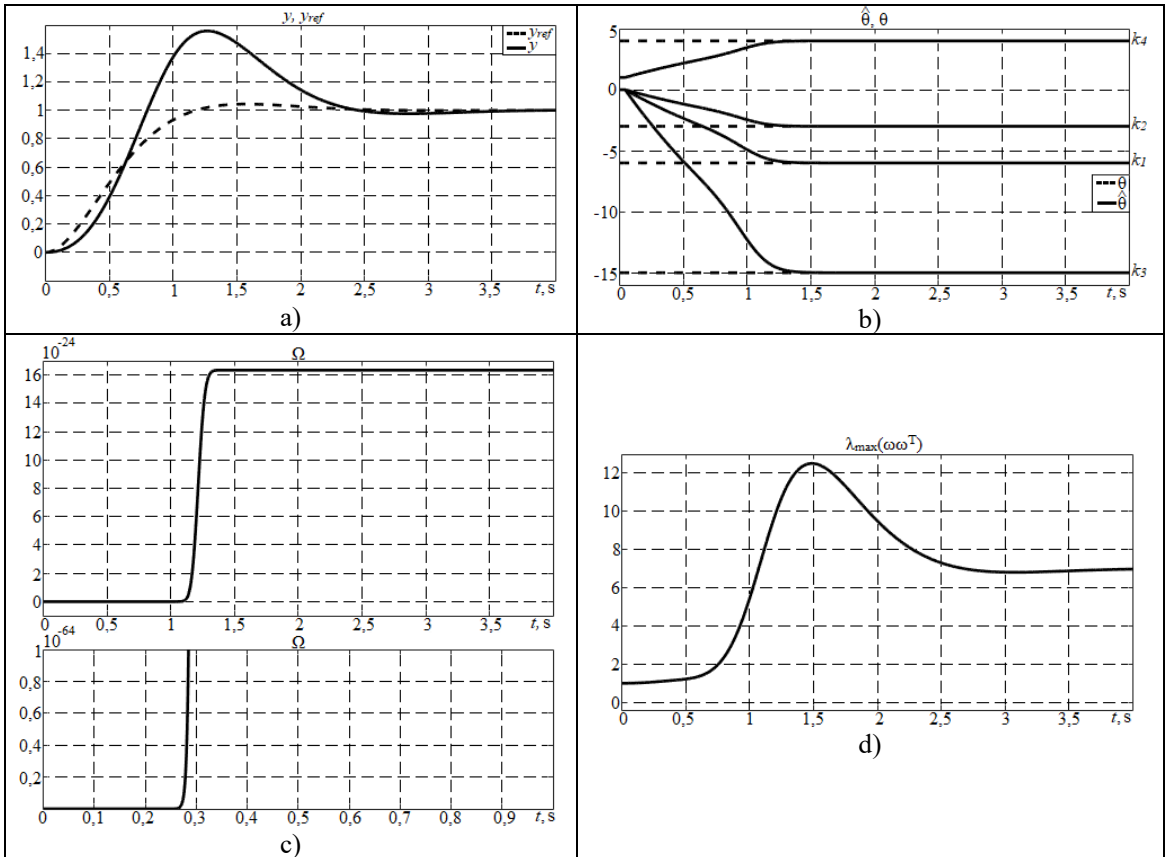

Figure 6. Transients of (a) plant $y$ and the reference model $y_{ref}$ outputs, (b) adjustable parameters $\hat{\theta}$ of control law, (c) regressor $\Omega$, and (d) value of $\lambda_{max}(\omega\omega^T)$

The simulation results validated the theoretical conclusions and confirmed the exponential stability of the augmented tracking error $\xi$ when the proposed adaptive law (2.2) was applied to solve the output feedback adaptive control problem.

Then, it was demonstrated that the adaptive law (2.2) was invariant to the sign of the gain $b_m$. For this purpose, the initial values of the control law parameters were set as follows: $\hat{\theta}(0) = [-1 \ 0 \ 0 \ 0]^T$. Figure 7 shows (a) the transients of $y$ and $y_{ref}$, and (b) the adjustable parameters $\hat{\theta}$ of the control law.

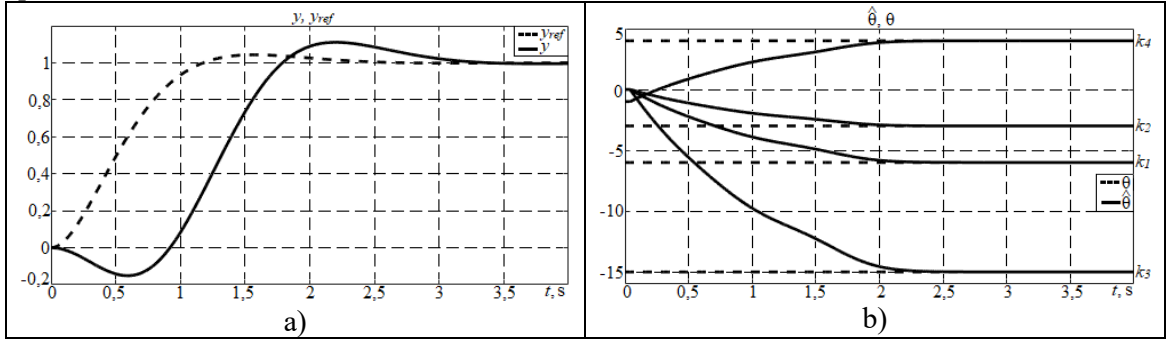

Figure 7. Transients of (a) $y$ and $y_{ref}$, (b) adjustable parameters $\hat{\theta}$ of control law

The transients shown in Figure 7 confirmed the correctness and efficiency of the adaptive law (2.2) under the condition that the sign of the gain $b_m$ was unknown.

Then the developed system was simulated in the practically important stabilization mode. For this purpose, the plant initial condition was set as $y(0) = 1$ and the reference signal was selected as $r = e^{-1t}$. The initial conditions of the reference model and other loop parameters were set according to (3.2.1) and (3.2.2).

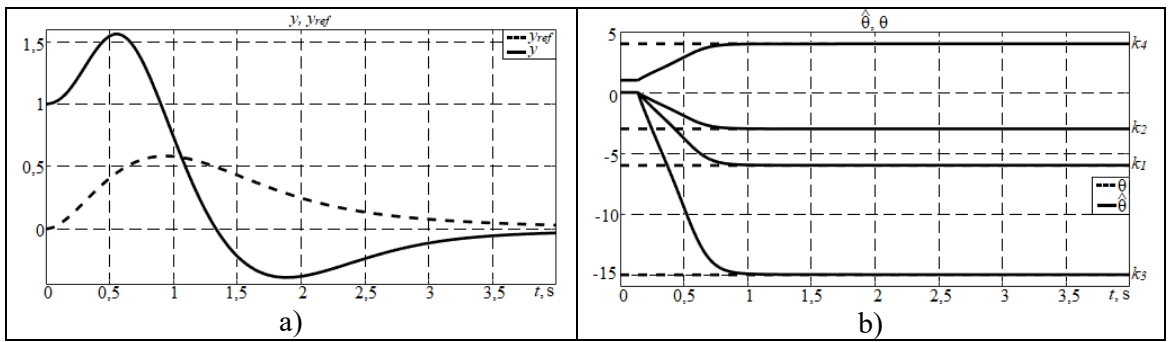

Figure 8. Transients of (a) $y$ and $y_{ref}$, (b) adjustable parameters $\hat{\theta}$ of control law

The simulation results, shown in Figure 8, demonstrated the effectiveness of the developed output feedback adaptive control system to solve the plant stabilization problem.

The next experiment was to compare the performance of the proposed output feedback system for different values of $\gamma_0$. Figure 9 shows the error norms $\|\xi\|$, which were obtained for different values of the parameter $\gamma_0$.

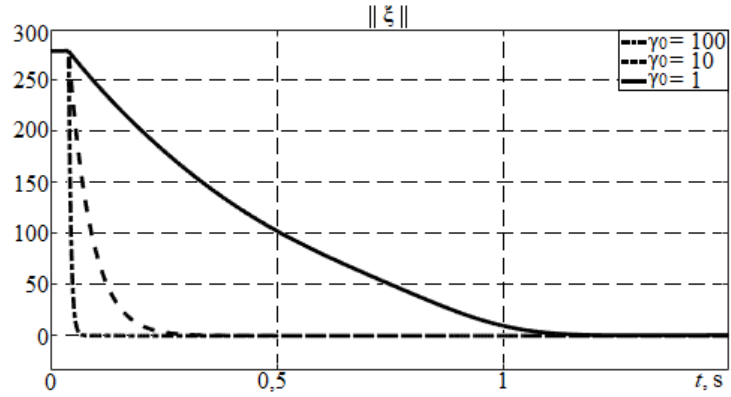
Figure 9. Dependence of $\|\xi\|$ from value of $\gamma_0$

The transients shown in Figure 9 confirmed the conclusions made in the proof of Theorem. They also demonstrated that the rate of the error $\xi$ convergence to zero could be improved if the value of $\gamma_0$ was made higher.

Finally, it was checked whether the developed output feedback adaptive control system allowed one to set the value of the minimal rate of the error $\xi$ convergence to zero with the help of $\gamma_1$ value choice, as it was noted in Remark 1. For this purpose, the developed system was modelled using the values $\gamma_1 = 0$, $\gamma_1 = 10$ and different reference signals $r$. Figure 10 presents (a) transients of the error norm $\|\xi\|$ when $\gamma_1 = 0$ and different $r$ were used, (b) $\gamma_1 = 10$ and the same set of $r$ were used.

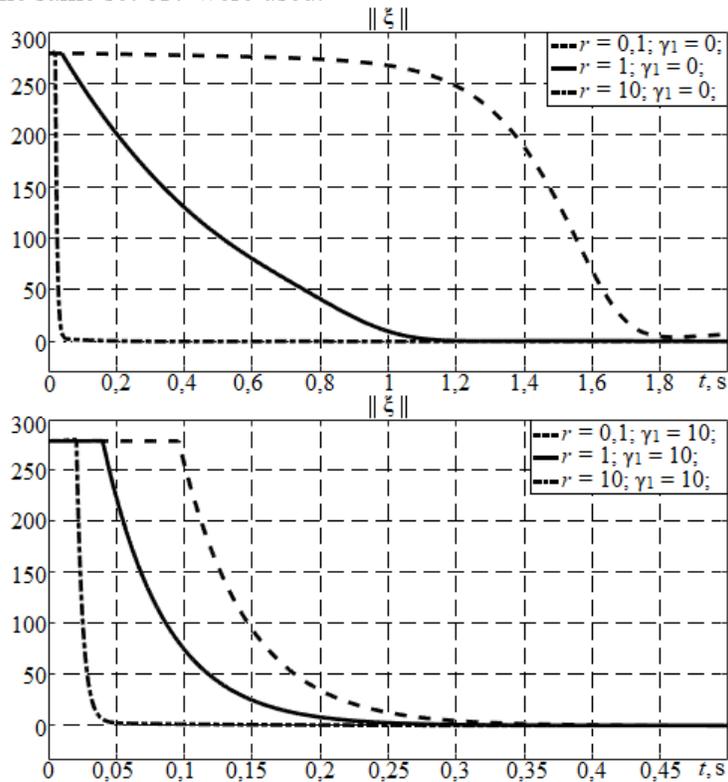
Figure 10. Dependence of $\|\xi\|$ from reference signal $r$ and value of $\gamma_1$

The results of the experiment shown in Figure 10 confirmed that the value of the minimal rate of the error $\|\xi\|$ convergence to zero could be adjusted by change of the $\gamma_1$ value.

## 4. Conclusion

The new adaptive law to adjust the controller parameters was proposed, which equally applicable to the linear error models with a measurable state (1.1) and output (1.2). Under the conditions that the regressor was finitely exciting $\Delta \in FE$ and the unknown controller ideal parameters $\theta$ were time-invariant, it provided exponential stability of the error $\xi$ and elementwise monotonicity of the controller adjustable parameters $\hat{\theta}$. In addition, it solved a number of problems of the adaptive control theory. In particular, the need to know the sign or values of the elements of the plant input gain vector $B$, the need for manual selection of the adaptive gain matrix $\Gamma$.

The scope of further research is to analyze the robustness of the developed adaptive control system to disturbances and formulate conditions, under which the requirement (1.6) is satisfied for the initial regressors of the parameterizations (2.1.9) and (2.2.16). Another aim is to extend these results to control plants with time-varying parameters. The second part of the paper will be devoted to the problem of state and output feedback adaptive control of linear plants with piecewise-constant unknown parameters, and the third part – linear plants with time-varying parameters of a certain type.

## **Appendix**

**Proof of theorem.** As $\Omega \in R$, the elementwise solution of (2.2) could be obtained:

$$\tilde{\theta}_i(t) = e^{-\int_{t_r^+}^{t} \gamma(\tau)\Omega^2(\tau)d\tau} \tilde{\theta}_i(t_r^+). \tag{A1}$$

As $\text{sign}(\gamma\Omega^2) = const > 0$, then it follows from (A1) that $|\tilde{\theta}_i(t_a)| \leq |\tilde{\theta}_i(t_b)| \ \forall t_a \geq t_b$, as was to be proved in the first part of the theorem. Let the second and third parts of the theorem be proved.

As the matrix $A_{ref}$ is a Hurwitz one in (1.1), then, according to the corollary of the Klaman-Yakubovich-Popov lemma, there always exist some scalar $d > 0$ and corresponding virtual directly unmeasurable output:

$$y_e = d\tilde{\theta}^T \omega, \tag{A2}$$

such that the transfer function $H(s) = d$ is strictly positive real, and the following equalities hold:

$$A_{ref}^T P + P A_{ref}^T = -qq^T - \mu P,$$
$$PB = \sqrt{2d}q. \tag{A3}$$

Then the following quadratic function is chosen to analyze the stability of the error equation (1.1) in case the adaptive law (2.2) is applied:

$$V = \xi^T H \xi = e_{ref}^T P e_{ref} + \frac{1}{2}\tilde{\theta}^T\tilde{\theta},$$

$$H = \text{blockdiag}\left\{P, \frac{1}{2}I\right\}, \underbrace{\lambda_{\min}(H)\|\xi\|^2}_{\lambda_m} \leq V(\|\xi\|) \leq \underbrace{\lambda_{\max}(H)\|\xi\|^2}_{\lambda_M}, \quad (A4)$$

where the matrix $P$ corresponds to one of the solutions of (A3).

The derivative of (A4) with respect to the equations (1.1) and (2.2) is written as:

$$\dot{V} = e_{ref}^T\left(A_{ref}^T P + P A_{ref}\right)e_{ref} + 2\tilde{\theta}^T \omega e_{ref}^T PB - \tilde{\theta}^T \gamma \Omega^2 \tilde{\theta} =$$
$$= -\mu e_{ref}^T P e_{ref} - e_{ref}^T q q^T e_{ref} + 2\tilde{\theta}^T \omega e_{ref}^T q\sqrt{2d} - \tilde{\theta}^T \gamma \Omega^2 \tilde{\theta}. \quad (A5)$$

Following (A2), as $d > 0$, then it is acceptable to consider $d = 0.5$ for certainty. In this case it is obtained from (A5):

$$\dot{V} = -\mu e_{ref}^T P e_{ref} - e_{ref}^T q q^T e_{ref} + 2\tilde{\theta}^T \omega e_{ref}^T q \pm \tilde{\theta}^T \omega \omega^T \tilde{\theta} - \tilde{\theta}^T \gamma \Omega^2 \tilde{\theta} =$$
$$= -\mu e_{ref}^T P e_{ref} - \left(e_{ref}^T q - \tilde{\theta}^T \omega\right)^2 + \tilde{\theta}^T \omega \omega^T \tilde{\theta} - \tilde{\theta}^T \gamma \Omega^2 \tilde{\theta} \leq \quad (A6)$$
$$\leq -\mu e_{ref}^T P e_{ref} + \tilde{\theta}^T \omega \omega^T \tilde{\theta} - \tilde{\theta}^T \gamma \Omega^2 \tilde{\theta}.$$

Let two different cases be considered: $t < t_e$ and $t \geq t_e$. As for the first one, according to (2.1) and (2.2), it holds that $\Omega = 0$ and $\|\tilde{\theta}\| = \|\tilde{\theta}(0)\|$. Then $\forall t < t_e$ the equation (A6) is rewritten as:

$$\dot{V} \leq -\mu e_{ref}^T P e_{ref} + \tilde{\theta}^T(0)\omega \omega^T \tilde{\theta}(0) \pm \tilde{\theta}^T \tilde{\theta} \leq$$
$$\leq -\mu e_{ref}^T P e_{ref} - \tilde{\theta}^T \tilde{\theta} + \tilde{\theta}^T(0)\omega\omega^T\tilde{\theta}(0) + \tilde{\theta}^T(0)\tilde{\theta}(0). \quad (A7)$$

Let the notion of the maximum eigenvalue of the matrix $\omega \omega^T$ over the time interval $[0; t_e)$ be introduced:

$$\delta = \sup_{\forall t < t_e} \max \lambda_{\max}\left(\omega \omega^T\right). \quad (A8)$$

Considering (A8), (A7) for $t < t_e$ is rewritten as:

$$\dot{V} \leq -\mu \lambda_{\min}(P)\|e_{ref}\|^2 - \|\tilde{\theta}\|^2 + (\delta+1)\|\tilde{\theta}(0)\|^2 \leq -\eta_1 V + r_B. \quad (A9)$$

where $\eta_1 = \min\left\{\frac{\mu \lambda_{\min}(P)}{\lambda_{\max}(P)}; 2\right\}; r_B = (\delta+1)\|\tilde{\theta}(0)\|^2$.

Having solved (A9), it is obtained:

$$\forall t < t_e: V \leq e^{-\eta_1 t}V(0) + \frac{r_B}{\eta_1}. \quad (A10)$$

Considering $\lambda_m \|\xi\|^2 \leq V$ and $V(0) \leq \lambda_M \|\xi(0)\|^2$, the estimate of $\xi$ for $\forall t < t_e$ is obtained from (A10):

$$\|\xi\| \leq \sqrt{\frac{\lambda_M}{\lambda_m} e^{-\eta_1 t}\|\xi(0)\|^2 + \frac{r_B}{\lambda_m \eta_1}} \leq \sqrt{\frac{\lambda_M}{\lambda_m}\|\xi(0)\|^2 + \frac{r_B}{\lambda_m \eta_1}}. \quad (A11)$$

It follows from this that $\xi$ is bounded $\forall t < t_e$.

Let the second case be studied. Considering the definition of $\gamma$ and the fact that $\forall t \geq t_e$ the inequality $0 < \Omega_{LB} \leq \Omega \leq \Omega_{UB}$ holds, it is obtained from (A6) $\forall t \geq t_e$:

$$\dot{V} \leq -\mu e_{ref}^{T} P e_{ref} + \tilde{\theta}^{T}\omega\omega^{T}\tilde{\theta} - \tilde{\theta}^{T}\frac{\left(\gamma_{0}\lambda_{max}\left(\omega\omega^{T}\right) + \gamma_{1}\right)\Omega^{2}}{\Omega^{2}}\tilde{\theta} = \quad (A12)$$
$$= -\mu e_{ref}^{T} P e_{ref} + \tilde{\theta}^{T}\omega\omega^{T}\tilde{\theta} - \tilde{\theta}^{T}\left[\gamma_{0}\lambda_{max}\left(\omega\omega^{T}\right) + \gamma_{1}\right]\tilde{\theta}.$$

The following inequality holds for $\forall \omega$:
$$\tilde{\theta}^{T}\omega\omega^{T}\tilde{\theta} - \tilde{\theta}^{T}\gamma_{0}\lambda_{max}\left(\omega\omega^{T}\right)\tilde{\theta} = \tilde{\theta}^{T}\underbrace{\left(\omega\omega^{T} - \gamma_{0}\lambda_{max}\left(\omega\omega^{T}\right)I\right)}_{\leq -\kappa I}\tilde{\theta} \leq 0. \quad (A13)$$

So (A12) is rewritten as:
$$\dot{V} \leq -\mu e_{ref}^{T} P e_{ref} - \tilde{\theta}^{T}\left(\kappa + \gamma_{1}\right)\tilde{\theta} \leq -\mu\lambda_{min}(P)\|e_{ref}\|^{2} - (\kappa + \gamma_{1})\|\tilde{\theta}\|^{2} \leq -\eta_{2}V, \quad (A14)$$

where $\eta_{2} = \min\left\{\frac{\mu\lambda_{min}(P)}{\lambda_{max}(P)}; 2(\kappa + \gamma_{1})\right\}$.

The inequality (A14) is solved, and it is obtained for $t \geq t_{e}$:
$$V \leq e^{-\eta_{2}t}V(t_{e}). \quad (A15)$$

Considering $\lambda_{m}\|\xi\|^{2} \leq V$, $V(t_{e}) \leq \lambda_{M}\|\xi(t_{e})\|^{2}$ and (A11), the estimate of $\xi$ for $t \geq t_{e}$ is obtained from (A15) as:
$$\|\xi\| \leq \sqrt{\frac{\lambda_{M}}{\lambda_{m}}e^{-\eta_{2}t}\|\xi(t_{e})\|^{2}} \leq \sqrt{\frac{\lambda_{M}}{\lambda_{m}}\left(\frac{\lambda_{M}}{\lambda_{m}}\|\xi(0)\|^{2} + \frac{r_{B}}{\lambda_{m}\eta_{1}}\right)}. \quad (A16)$$

Therefore, together with (A1), it follows that $\xi \in L_{\infty}$ and $\xi$ converges to zero exponentially for all $t \geq t_{e}$ at a rate, which is directly proportional to the parameters $\gamma_{0}$, $\gamma_{1}$, which were to be proved in the second and third parts of Theorem.

**Proof of Proposition 2.** To prove the first part of the proposition, let $|\Delta| \leq c_{1}e^{c_{2}t}$ be considered, and $\Delta^{2} \leq c_{1}^{2}e^{2c_{2}t}$ be substituted into the definition of the regressor $\Omega$:
$$\Omega(t) = \int_{t_{r}^{+}}^{t} e^{-\sigma\tau}\Delta^{2}(\tau)d\tau \leq c_{1}^{2}\int_{t_{r}^{+}}^{t} e^{(2c_{2}-\sigma)\tau}d\tau. \quad (A17)$$

Considering $\sigma > 2c_{2}$, it is obtained from (A17) that:
$$\Omega(t) \leq c_{1}^{2}\int_{t_{r}^{+}}^{t} e^{-c_{3}\tau}d\tau = \frac{c_{1}^{2}}{c_{3}}\left(1 - e^{-c_{3}t}\right) \leq \frac{c_{1}^{2}}{c_{3}}. \quad (A18)$$

Therefore, $\Omega(t) \in L_{\infty}$ $\forall t \geq t_{r}^{+}$, which was to be proved in the first part of the proposition.

Let the second part of the proposition be proved. Following (1.6), the finite excitation condition is written for the regressor $\Delta$ over $\left[t_{r}^{+}; t_{e}\right]$:
$$\int_{t_{r}^{+}}^{t_{e}} \Delta^{2}(\tau)d\tau \geq \alpha. \quad (A19)$$

Then $\forall t \geq t_{e}$ the following inequality holds:
$$\int_{t_{r}^{+}}^{t} \Delta^{2}(\tau)d\tau > 0. \quad (A20)$$

As $\forall t < t_e \; e^{-\sigma t} > 0$ and (A20) holds, then the following also holds:

$$\Omega(t) = \int_{t_r^+}^{t} e^{-\sigma\tau}\Delta^2(\tau)d\tau > 0, \; \forall t \geq t_e. \tag{A21}$$

For further proof, let the following notation be introduced on the basis of the definition of $\Omega$:

$$\Omega(t) = \int_{t_r^+}^{t_e} e^{-\sigma\tau}\Delta^2(\tau)d\tau + \int_{t_e}^{t} e^{-\sigma\tau}\Delta^2(\tau)d\tau. \tag{A22}$$

Considering $\Delta^2 \leq c_1^2 e^{2c_2 t}$, the first integral of (A22) is bounded:

$$\int_{t_r^+}^{t_e} e^{-\sigma\tau}\Delta^2(\tau)d\tau \leq c_1^2 \int_{t_r^+}^{t_e} e^{-c_3\tau}d\tau = \frac{c_1^2}{c_3}\left(e^{-c_3 t_r^+} - e^{-c_3 t_e}\right). \tag{A23}$$

Then, considering (A18), (A22) and (A23), $\forall t \geq t_e$ it holds that:

$$0 < \frac{c_1^2}{c_3}\left(e^{-c_3 t_r^+} - e^{-c_3 t_e}\right) \leq \Omega(t) \leq \frac{c_1^2}{c_3}, \tag{A24}$$

which completes the proof of Proposition 2.